\documentclass[twocolumn,showpacs,amsmath,amssymb,prb,superscriptaddress]{revtex4}
\usepackage{graphicx}
\usepackage{bm}
\begin{document}
\title{Inhomogeneous states in two-dimensional frustrated phase separation.}

\author{C.  Ortix}
\affiliation{Insitute-Lorentz for Theoretical Physics, Universiteit Leiden, P.O. Box 9506, 2300 RA Leiden, The Netherlands. }
\author{J.  Lorenzana} 
\affiliation{SMC-INFM-CNR and Dipartimento di Fisica, Universit\`a di Roma 
``La Sapienza'', P.  Aldo Moro 2, 00185 Roma, Italy.}
\affiliation{ISC-CNR, Via dei Taurini 19, 00185 Roma, Italy.} 
\author{C.  Di Castro} 
\affiliation{SMC-INFM-CNR and Dipartimento di Fisica, Universit\`a di Roma 
``La Sapienza'', P.  Aldo Moro 2, 00185 Roma, Italy.}
\date{\today}

\begin{abstract}
We derive the phase diagram of a paradigmatic model of Coulomb frustrated phase separation in two-dimensional systems with  negative short-range electronic compressibility. We consider the system subject either to the truly three-dimensional long-range Coulomb interaction (LRC) and to a two-dimensional LRC with logarithmic-like behavior. In both cases we find that the transition from the homogeneous phase to the inhomogeneous phase is generically first-order except for a critical point. Close to the critical point, inhomogeneities arrange in a triangular lattice with a subsequent first-order topological transition to stripe-like objects by lowering the Coulomb frustration. A proliferation of inhomogeneities which have inside smaller inhomogeneities is expected near all the transition lines in systems embedded in the three-dimensional LRC alone.  
\end{abstract}


\maketitle

\begin{section}{Introduction}
\label{sec:intro}
Domain pattern formation is a beautiful example of cooperative
behavior in complex systems with competing interactions on different
length scales \cite{seu95}. In electronic systems, this idea has
gained momentum due to theoretical and experimental studies in
materials like cuprates and manganites. Indeed it has become clear
that strong electron correlations generally produce a tendency towards
phase separation in electron-rich and electron-poor regions
\cite{nag83,mar90,eme90c,lor91,cas92,low94,cas95b,nag98}. Although the long-range part of the
Coulomb interaction (LRC) spoils phase separation (PS) as a
thermodynamic phenomenon, the frustrated tendency towards charge
segregation is still important and gives rise to inhomogeneous states
where the charge is segregated over some characteristic distance but
the average density is a fixed constant in order to avoid a diverging
Coulomb cost in the thermodynamic limit
\cite{nag83,mar90,low94,cas95b,nag98,lor01II,lor02,ort06,ort07,ort07b,ort08}.   

From the experimental point of view, a prominent emergent tool in
recent years has been scanning tunneling microscopy (STM). 
In the high temperature superconductors context, STM experiments in
cuprates \cite{lan02} have revealed a phase segregation on the scale
of $\sim 20$ lattice constants reconcilable with a Coulomb frustrated
phase separation (FPS) between an underdoped pseudogap phase and a
superconducting phase at higher dopings.  
Coexistence of insulating and metallic regions on the scale of tens to
thousands of nanometers have been reported also in thin films of
colossal magnetoresistance manganites \cite{bec02}. Noticeably
percolation of the metallic regions is closely correlated to abrupt
changes in transport suggesting that FPS is at the heart of the
colossal magnetoresistance behavior \cite{bec02,zha02}. 
More microscopic configurations with stripe patterns have been
observed in cuprates\cite{tra95}, nickelates\cite{tra94} and manganites\cite{mor98}.   
Even more, the discovery of pronounced anisotropies in  transport
measurements in ruthenates \cite{bor07} and in GaAs heterostructures
at weak magnetic fields \cite{lil99} are in line with the proposal of
exotic electronic liquid phases \cite{kiv98} analogue to the
intermediate order phases of liquid  crystals \cite{cha95}.  Evidence
for mesoscopic electronic phase separation have been recently
presented in organics\cite{col06,col08},
graphene\cite{mar08} and pnictides\cite{par09}.

The generality of this phenomena calls for simple models
which neglect the microscopic details of the specific system rather
capturing its general properties. The tendency towards PS is then
caused by the appearance of anomalies in the electronic contribution
to the free energy of the system $f_e$,  as a function of the density $n$.  Two kind
of anomalies have been identified as the most relevant ones for phase
separation in  
electronic systems \cite{fin07,ort07b,ort08}. The first anomaly
consists of a range of densities where the compressibility is negative
while  the second one corresponds to a single point where the inverse of the electronic
compressibility has a Dirac-delta-like negative divergence due to a
 crossing of the free energies of the two competing phases\cite{ort07b,ort08,jam05}. The two anomalies 
can be labeled by the exponent $\gamma$ characterizing the behavior of
the electronic free energy around  
a reference  density $n_c$, i.e. $f_e = \alpha |n-n_{c}|^\gamma$ with $\alpha<0$ and $\gamma=2$ (negative compressibility region) or  $\gamma=1$ (cusp
behavior). Both anomalies appear often in model computations of electronic
systems, for example  Refs.~\onlinecite{nag98,cas95b,kag99,lor01II}
correspond to $\gamma=2$ while Refs.~\onlinecite{bri99b,oka00} to $\gamma=1$.

  When LRC effects can be considered as a weak perturbation upon the
  PS mechanism, one can achieve a universal picture of the FPS
  phenomenon. Antithetically, upon strong frustrating effects, the two
  short-range compressibility anomalies give rise to two different
  universality classes. Within the $\gamma=1$ universality class,
  different studies in 3D systems \cite{lor01I,lor02} and in 2D
  systems \cite{ort06,ort07,jam05} have enlightened the key role of
  the system dimensionality in the FPS mechanism. Recently, the full
  phase diagram for the $\gamma=2$ class has been characterized in
  Ref.~\onlinecite{ort07bis} both in isotropic and strongly anisotropic  3D
  systems. The aim of this work is to investigate the properties of
  the FPS phase diagram for $\gamma=2$ at strong frustration in 2D
  systems. This is an important generalization as many interesting strongly correlated
  systems are layered as cuprates, nickelates, iron oxypnictides superconductors, or
  are truly two dimensional as the two-dimensional electron 
  gas in heterostructures, graphene, etc. Relevant for cuprates,
  manganites and nickelates is how charge density wave instabilities
  of a uniform phase\cite{cas95b} evolve into the very anharmonic
  structures experimentally observed\cite{tra95,tra94,mor98}. 

We will consider two
  versions of the Coulomb interaction labeled by an effective
  dimensionality $d$. The $d=3$ case corresponds to the usual three
  dimensional Coulomb interaction and applies to the physical
  systems above mentioned  (layered systems,
  heterostructures, graphene, etc.). The    
 $d=2$ case, instead, corresponds to a 
  fictitious logarithmic Coulomb interaction.\cite{mur02}  The
  effective dimensionality $d$ regards only the interaction
  and should not be confused with the dimensionality of the system
  which is 2D. Alternatively the 2D system with logarithmic
  interaction can be considered as an anisotropic 3D system subject to
  the conventional Coulomb interaction where modulations in one ``hard'' direction  are forbidden. 
  This anisotropy can originate in the underling crystal structure and 
  the case of only one hard direction (as opposed to two hard
  directions considered in Ref.~\onlinecite{ort07bis}) corresponds to 
  the case of weakly coupled chains. This is relevant for electronic
  phase separation in organics where elongated domains have 
  been reported.\cite{col06,col08}

\end{section}

\begin{section}{Phase diagrams in 2D systems}
\label{sec:model}

Our starting point is a paradigmatic model of Coulomb frustrated phase separation in systems with a short-range negative compressibility density region that is defined by the following free energy:
\begin{eqnarray}
F&=&\int d{\bf x} \,\,c \left| \nabla n({\bf x}) \right|^2+ \alpha \left[n({\bf x})-n_c\right]^2+ \beta \left[n({\bf x})-n_c\right]^4 \nonumber \\
& + & \dfrac{e^2}{2 \varepsilon_0}  \int d{\bf x} \int d{\bf x^{\prime}}\dfrac{\left[n({\bf x})-\overline{n}\right] \left[n({\bf x^{\prime}})-\overline{n}\right]}{G^{-1}(|{\bf x-x^{\prime}}|)} 
\label{eq:modeldim}
\end{eqnarray}
Here the gradient term models the surface energy of smooth interfaces and is parametrized by the stiffness constant $c$ whereas $\alpha<0$ is proportional to the inverse short-range compressibility. The inclusion of a fourth order term with $\beta>0$ provides a symmetric double-well form of the short-range part of the free energy with minima at $n=n_c\pm [|\alpha|/(2 \beta)]^{1/2}$.  
In general, $\alpha$ will depend on external parameters like the pressure. It can be also taken as temperature-dependent, as in Landau theory, in which case the model becomes a mean-field description of a temperature-driven transition to an inhomogeneous state. 
In addition, $\varepsilon_0$ indicates a static dielectric constant due to external degrees of freedom and $\overline{n}$ is the average density. A rigid background ensures charge neutrality. Finally $G(|{\bf x-x^{\prime}}|)$ is a positive-definite LRC kernel that in the $d=2$ case has a logarithmic-like behavior whereas for $d=3$, $G^{-1}(|{\bf x-x^{\prime}}|)=|{\bf x-x^{\prime}}|$. 

Since the model Eq.~(\ref{eq:modeldim}) has several parameters, it is
convenient, in order to study the phase diagram, to measure all
lengths in unit of the bare correlation  length $\xi=\sqrt{2 c/
  \alpha}$, densities in units of  $[\alpha/(2 \beta)]^{1/2}$ and
energy densities in terms of the barrier height $f_0=\alpha^2/(4
\beta)$. Then one reaches a dimensionless form consisting, apart from
an irrelevant constant, of a $\phi^4$ model augmented with a long-range Coulomb interaction. The corresponding Hamiltonian is defined by:  
\begin{eqnarray}
  {\cal H}&=&\int d{\bf x} \left[\phi^{2}({\bf x})-1\right]^2+|\nabla \phi({\bf x})|^2+\frac{Q^2}{2} \nonumber \\  & \times &  \int d{\bf x} \int d{\bf x^{\prime}}\dfrac{\left[\phi({\bf x})-\overline{\phi}\right] \left[\phi({\bf x^{\prime}})-\overline{\phi}\right]}{G^{-1}(|{\bf x-x^{\prime}}|)} 
\label{eq:model}
\end{eqnarray}
where, $Q^2$ is dimensionless renormalized frustration parameter of the long-range interaction given by
$$Q^2=\dfrac{e^2}{\varepsilon_0}\dfrac{\xi^{4-d}}{|\alpha|},$$
 and the classical scalar field $\phi$ represents the dimensionless
 local charge density with average density $\overline{\phi}$. 
The effect of frustration can also be measured by the parameter $Q^{2/(5-d)}$ introduced in Refs. \onlinecite{lor01I},\onlinecite{ort06},\onlinecite{ort07}. It can be obtained from the following dimensional analysis. Taking $\phi \sim 1$, the Coulomb energy density  per domain can be estimated making the integrals in a volume of order $l_d^2$ as $Q^2 l_d^{4-d}$ with $l_d$ indicating the typical size of the inhomogeneities measured in unit of $\xi$. Within a sharp interfaces approach, the surface energy density goes like $1 \,/ \,l_d$. Both quantities are optimized whenever $l_d \sim 1 \, / \, Q^{2 / (5-d)}$. Then $Q^{2 / (5-d)}$ takes the meaning of ratio between the additional energy cost induced by frustration and the typical PS energy gain.

For both the truly long-range Coulomb interaction and the logarithmic-like interaction, the charge susceptibility in momentum space can be derived  by computing the static response to an external field. At ${\bf k} \neq 0$ one finds:
\begin{equation}
\chi({\bf k},d)=\left[{\bf k}^2+\dfrac{2^{3-d} \pi Q^{2}}{\left|{\bf k}\right|^{4-d}}-2+6 \overline{\phi}^{2}\right]^{-1}. 
\end{equation}
Notice that the second term in the brackets yields the familiar forms
of the Fourier transform of the Coulomb interaction for $d=2,3$. 

The charge susceptibility has a peak at finite momentum $k_0=\left[(4-d)2^{2-d}\pi Q^{2}\right]^{1/(6-d)}$ determined by the competition between interface and charging effects. It diverges as $Q$ approaches a Gaussian instability line :
\begin{equation}
Q_g=Q_c\left[1-3 \overline{\phi}^{2}\right]^{(6-d)/4}
\label{eq:gaussinstability}
\end{equation} 
where
$$Q_c=\left[\dfrac{2}{1+\frac{2}{(4-d)}}\right]^{(6-d)/4} \left[2^{2-d} \pi (4-d)\right]^{-1/2}$$
represents the maximum frustration degree above which the system is in its homogeneous phase. 

The Gaussain instability line Eq.~(\ref{eq:gaussinstability}) indicates an instability towards a sinusoidal charge density wave (SCDW) of period $2 \pi /k_0$ and direction chosen by spontaneous symmetry breaking. Still, it cannot persist up to $Q \rightarrow 0$ since inhomogeneities are predicted only for average densities inside the ``miscibility'' gap \cite{fin07}  $|\overline{\phi}|<1/\sqrt{3}$. This is not in tune with the usual Maxwell construction that on the contrary predicts  an inhomogeneous state for $|\overline \phi|<1$. 
Analogously to 3D systems \cite{ort07bis} indeed, we find that the inclusion of non-gaussian terms results in a first-order transition preempting the Gaussian
instability except for the critical point (CP)
$(\overline{\phi},Q)=(0,Q_c)$ as we show below.

\begin{figure}[tbp]
\includegraphics[width=6.5cm]{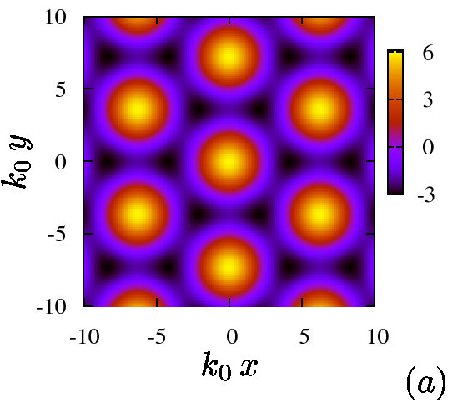} \\
\includegraphics[width=6.5cm]{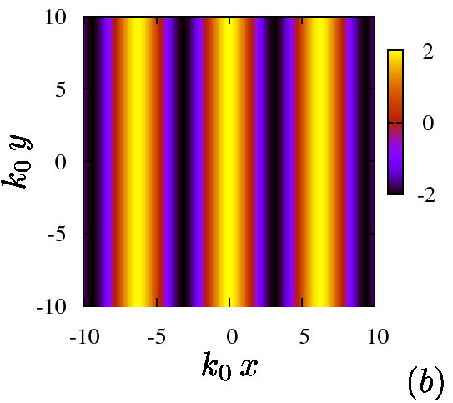}
\caption{(Color online) (a) Contour plot of the charge density modulation $\left[\phi({\bf x})-\overline{\phi}\right]/\phi_G$ near the critical point for a triangular crystal of inhomogeneities. The yellow spots are in the vicinity of the triangular lattice points where $\phi({\bf x})=\overline{\phi}+6 \phi_G$. (b) Same for a unidirectional SCDW. 
In both panels, lengths have been measured in units of $1/k_0$ for convenience. }
\label{fig:contourharmonic}
\end{figure}

By restricting to periodic textures, the free energy density difference between a modulated phase and the homogeneous phase can be expressed in momentum space as:
\begin{eqnarray}
\dfrac{\delta F}{V}&=&\sum_{{\bf G}\neq 0} \phi_{{\bf G}} \chi^{-1}({\bf G},d) \phi_{-{\bf G}}+ \label{eq:fourier}\\ & &+4 \overline{\phi} \sum_{{\bf G}_1,{\bf G}_2,{\bf G}_3\neq 0} \phi_{{\bf G}_1} \phi_{{\bf G}_2} \phi_{{\bf G}_3} \delta_{{\bf G}_1+{\bf G}_2+{\bf G}_3,0}+\nonumber \\ & &
+\sum_{{\bf G}_1,{\bf G}_2,{\bf G}_3,{\bf G}_4\neq 0} \phi_{{\bf G}_1} \phi_{{\bf G}_2} \phi_{{\bf G}_3} \phi_{{\bf G}_4} \delta_{{\bf G}_1+{\bf G}_2+{\bf G}_3+{\bf G}_4,0} \nonumber
\end{eqnarray}
where the ${\bf G}$'s are the wavevectors of the reciprocal lattice and $V$ indicates its unit cell volume. 
The presence of the cubic term in Eq.~(\ref{eq:fourier}) opens the way to first-order transitions except for the CP. Away but close to the CP, the transition will be weakly first-order. Hence it can be treated, analogously to the liquid-solid transition\cite{ale78,cha95}, by fixing the wavevectors to have an equal magnitude $\left|{\bf G}\right|=k_0$.  In this case, to gain an energetic advantage from the cubic term, one needs triads of wavevectors forming an equilateral triangle so that the Kronecker delta is satisfied. In 2D systems this is only verified by the hexagonal reciprocal lattice that is defined by the set of six wavevectors ${\bf G}/k_0=(\pm \cos{\pi/3},\pm \sin{\pi/3}),(\pm 1,0)$. The corresponding free energy density reads:
\begin{equation}
\dfrac{\delta F}{V}=\left[-2+ 6 \overline{\phi}^2+ 2 \left(\dfrac{Q}{Q_c}\right)^{\frac{4}{(6-d)}}\right] 6 \phi_G^2 + 48 \overline{\phi} \phi_G^3+90 \phi_G^4
\label{eq:deltaftri}
\end{equation} 
As a result, one finds that the first structure to become stable corresponds to droplet-like structures
forming a triangular crystal of inhomogeneities. 
Fig.~\ref{fig:contourharmonic} (a) shows the corresponding charge modulation close to
the critical point.

Upon minimizing Eq.~(\ref{eq:deltaftri})  with respect to the wave
amplitude $\phi_{G}$ and requiring $\delta F / V =0$, one finds that
the first-order transition line is given by:  
\begin{equation*}
Q_{T}=Q_c\left[1-\dfrac{37}{15} \,\overline{\phi}^{\,2}\right]^{(6-d)/4}.
\end{equation*}

On entering in the inhomogeneous phase density region, the triangular
crystal phase is expected to compete with a unidirectional SCDW in
which the local charge is modulated along stripes [see Fig.~\ref{fig:contourharmonic}(b)] whose free energy density reads:
\begin{equation}
\dfrac{\delta F}{V}=\left[-2+ 6 \overline{\phi}^2+ 2 \left(\dfrac{Q}{Q_c}\right)^{\frac{4}{(6-d)}}\right] 2 \phi_G^2 + 6  \phi_G^4
\label{eq:deltafscdw}
\end{equation} 

We
find that at fixed frustration $Q$, stripes become stable close to
$\overline{\phi}=0$ [see Fig.~\ref{fig:topo}]. This leads to a
first-order morphological transition that partially restores the
translation symmetry. In the frustration-density
plane, the morphological transition line can be determined by equating the free energies densities Eqs.~(\ref{eq:deltaftri}),~(\ref{eq:deltafscdw}) of the two phases minimized with respect to the wave amplitude $\phi_G$. We find : 
\begin{equation*}
Q_M=Q_c\left[1-\dfrac{87}{19-6\sqrt{6}} \overline{\phi}^2\right]^{(6-d)/4}.
\end{equation*}

\begin{figure}[tbp]
\includegraphics[width=7cm]{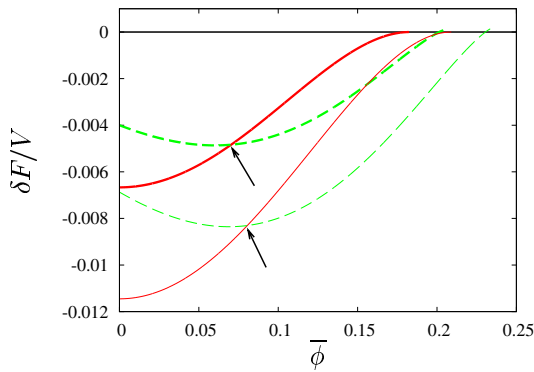}
\caption{(Color online) Comparison for $Q=0.9\, Q_c$ of the free energy densities of
  the triangular crystal phase (full lines) and the SCDW (dashed
  lines) for a $d=2$ LRC (thick lines) and the truly $d=3$ LRC (thin
  lines). We have substracted the free energy density of the
  homogeneous phase (horizontal line). Arrows indicate the cusp singularity
  induced in both cases at the topological transition.}
\label{fig:topo}
\end{figure}

Although the transition lines are only asymptotically exact close to
CP \cite{ort07bis}, we find the topology of the phase diagram found in
the strong frustration regime to persist at weak frustration. This is shown
in Fig.~\ref{fig:phasediagram} for a $d=2$ Coulomb interaction where
we characterized the full phase diagram. In the weak frustration regime ($Q<<1$),
we considered a Uniform Density Approximation (UDA) where we assumed
domains of uniform density of one or the other phases separated by
sharp interfaces \cite{nus99,lor01I,lor02,mur02,ort07b}. The UDA is an accurate
description in this regime since there is a sufficient separation among
the typical size of the domains
$l_d$ 
and the screening length $l_s$ that controls the charge relaxation inside
the domains \cite{ort07b}.   
For $d=2,3$, indeed, the latter goes like \cite{ort07b}:
$$l_s \sim \dfrac{1}{Q^{\frac{2}{(4-d)}}}$$
In addition, $l_d$ is much larger then the typical Gaussian instability wavelength $1/k_0$. Thus one can rely on a sharp interface approach where surface energy effects are taken into account by neglecting long-range effects and computing the excess energy of an isolated interface. Then one recovers the surface tension of a $\phi^4$ kink \cite{cha95}
$\sigma=8/3$. 

\begin{figure}[tb]
\includegraphics[width=7cm]{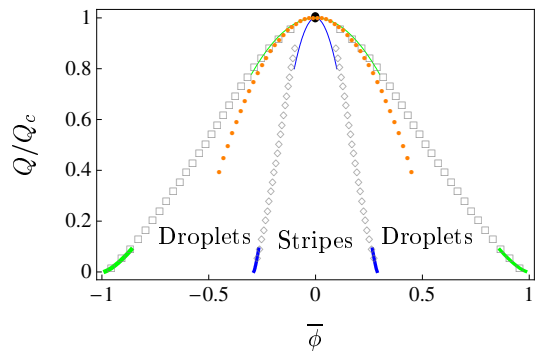}
\caption{(Color online) Phase diagram in two-dimensional systems embedded in the
  $d=2$ long-range interaction. The small dots indicates the Gaussian
  instability line $Q_g$. The thin (thick) lines represent first-order
  transitions in the strong (weak) frustration approximation. In the two
  limits they overlap with the corresponding numerically determined
  transition lines from the homogeneous phase to droplet
  inhomogeneities ($\square$) and from droplets to stripes
  ($\diamondsuit$). Finally the black circle indicates the CP. } 
\label{fig:phasediagram}
\end{figure}

In the weak frustration regime, we find the same topological transitions (thick lines in Fig.~\ref{fig:phasediagram}) as
in strong frustration but now the inhomogeneities form sharply defined
circular droplets and stripes. We have also studied 
the crossover from weak to strong frustration minimizing a discretized
version of the model 
Eq.~(\ref{eq:model}) in the Wigner-Seitz Approximation. The corresponding
transition lines are shown with squares and diamonds in
Fig.~\ref{fig:phasediagram}. 
For droplet-like structure we assumed circular symmetry in order to reduce the effective dimensinality of the minimization procedure to one. 
The upper panel of Fig.~\ref{fig:densprofile} shows the corresponding charge density
profile in the radial direction at $\overline{\phi}=0.5$ and different
frustrations. By decreasing the frustration degree, the charge density modulation in the radial direction gets more unharmonic similarly to what we find for striped states at $\overline{\phi}=0$ [see lower panel of Fig.~\ref{fig:densprofile}].  

\begin{figure}[tb]
\includegraphics[width=7cm]{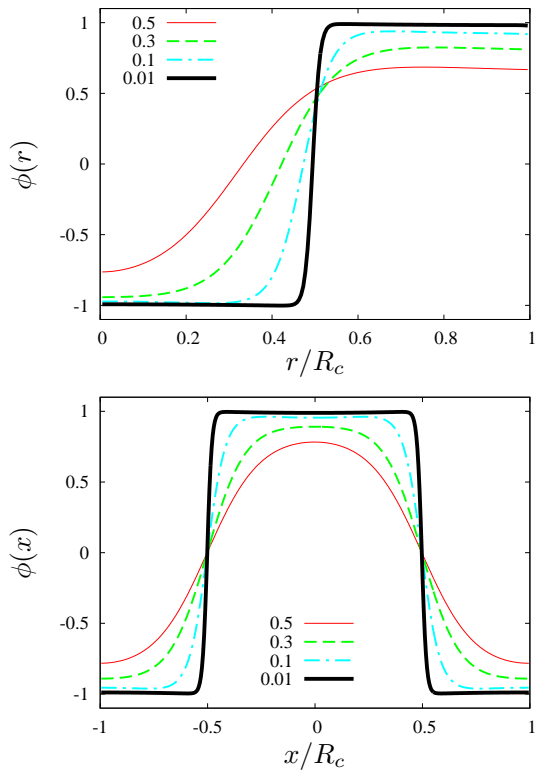}
\caption{(Color online). Top panel:
Behavior of the local charge density modulation for droplets at $\overline{\phi}=0.5$ and different values of the Coulomb frustration $Q/Q_c$. As the Coulomb frustration decreases, unharmonicity is build in and the charge density profile tends to a square wave in the radial direction $r$ (thick full line). Bottom panel: Same for $\overline{\phi}=0$ where only stripe-like objects appear.} 
\label{fig:densprofile}
\end{figure}

The features of the FPS phase diagram are dramatically different if the stiffness constant is made strongly anisotropic. More in detail, we assume that $c_{\parallel} / c_{\perp}>1$ where $c_{\parallel}$ ($c_{\perp}$) indicates the stiffness component in the ``hard'' (``soft'') direction of the 2D plane and focus on the limit where any charge modulation in the ``hard'' direction is completely forbidden, {\it i.e.} $c_{\parallel} / c_{\perp} \rightarrow \infty$. 
In a 2D system subject to the logarithmic-like $d=2$ interaction, in
view of the analogy discussed in Sec.~\ref{sec:intro}, this would
correspond to consider a 3D system with two ``hard'' directions. The
corresponding phase diagram has been shown in Fig.~3 of
Ref.~\onlinecite{ort07bis}. In that case it has  been shown that this
strong anisotropy allows for  
 second- and first-order transitions transition lines joined by a tricritical point. 

We now show that this feature appears also in 2D systems subject to the ordinary $d=3$ Coulomb interaction. 
Since the strong anisotropy allows only for unidirectional charge modulations, the cubic term of Eq.~(\ref{eq:fourier}) identically vanishes by restricting to wavevectors of $k_0$ magnitude. Thus, to include non Gaussian terms, one has to retain at least two collinear wavevectors so that the Fourier decomposition of the order parameter reads:
$$\phi({\bf x})=\overline{\phi}+2 \phi_1 \cos{(G_1\cdot x)}+2 \phi_2 \cos{(G_2\cdot x)}$$
where $|G_2|=2 |G_1|=2 k_0$. By minimizing with respect to the periodicity of the charge density modulation and assuming $\phi_2<<\phi_1$, that can be checked {\it a posteriori}, one obtains the following Landau free energy expansion in terms of $\phi_1$:
\begin{equation}
\dfrac{\delta F_{II}}{V}=r \phi_1^2+u_4 \phi_1^4+u_6 \phi_1^6
\label{eq:enertricrit}
\end{equation} 
where the quadratic term coefficient $$r=4 \left[\left(\dfrac{Q}{Q_c}\right)^{4/3}-\left(\dfrac{Q_g}{Q_c}\right)^{4/3}\right]$$ vanishes along the Gaussian  instability line whereas
 $u_4=6-54 \overline{\phi}^2 (Q_c/Q)^{4/3}$ and $u_6$ is a positive
 constant. It is simple to notice that this is the canonical form of
 the Landau free energy expansion around a tricritical point that is
 determined by the vanishing of $u_4$ along the Gaussian instability
 \cite{cha95} and is therefore given by $(\overline{\phi},Q)=(1/(2
 \sqrt{3}), (3/4)^{3/4})$.  
The finite density window of the second order line is important
because it allows for a second order quantum critical point (QCP) as
discussed in Ref.\onlinecite{cas95b}. Furthermore as parameters are changed
(density, frustration) an evolution similar to the one displayed in
Fig.~\ref{fig:densprofile} will occur. This establishes a 
connection between the Gaussian instabilities predicted in models of
the cuprates\cite{cas95b} and the probably anharmonic stripe patterns
observed\cite{tra95}.

Apart from  small shifts of the
  transition lines, the effective dimensionality of the Coulomb interaction does not change the features of the 2D phase diagram both in absence and in presence of strong anisotropies. 
This is reported in Fig.~\ref{fig:anisotropicphaseidiagram} where we show the phase diagram of an anisotropic 2D system subject to the truly $d=3$ Coulomb interaction. 
An important difference, however, will appear close to the
  transition lines as discussed below.

\begin{figure}[tb]
\includegraphics[width=7cm]{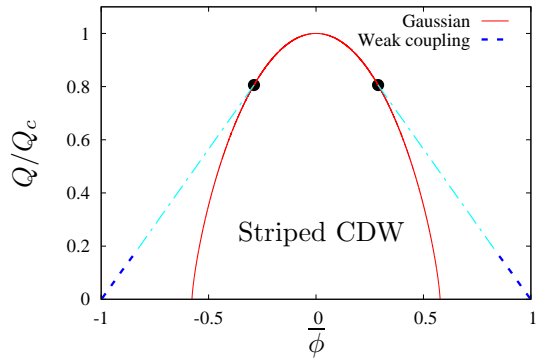}
\caption{(Color online).The phase diagram for anisotropic 2D systems subject to the truly $d=3$ Coulomb interaction. The thin line corresponds to the Gaussian instability that determines the second-order transition line above the tricritical point (black circle). Below the tricritical point, the transition is first-order and is determined in the weak coupling regime assuming the UDA (dashed lines). The long-short dashed lines are qualitative lines of the crossover from the weak to the strong coupling regime.} 
\label{fig:anisotropicphaseidiagram}
\end{figure}

\end{section}

\begin{section}{Discussion and conclusions}
The topology of the FPS phase diagrams in 2D systems has a strong
similarity to the 3D case of Ref.~\onlinecite{ort07bis} except that a
crystal of drops appears in addition in 3D. Thus one can
state that the system dimensionality $D$ and the effective
dimensionality of the long-range Coulomb interaction $d$ do not
qualitatively affect the FPS phenomenology in the $\gamma=2$
universality class.   This is not quite the end of the story since
the transition from the 
homogeneous to the inhomogeneous phase and the topological transitions
 are first-order-like. A cusp singularity in the minimal free
energy of the system will be naturally produced as indicated by arrows
in Fig.~\ref{fig:topo}. 
This drives the
system towards a new stage of FPS tendency  which now, however, is
governed by the $\gamma=1$ universality class and a new renormalized
frustration constant $Q'$ which depends on the distance from the critical
point and diverges at the critical point.  
At this second stage of
domain pattern formation, a dramatic difference between 3D and 2D
$d=3$ systems will arise. Indeed in 3D systems, for $\gamma=1$, 
there is a critical
value of $Q'$ above which the second stage of pattern formation is
suppressed\cite{lor01I,lor02,ort07,ort07b} while in a 2D, $d=3$ system the second
stage always occurs no matter 
how big is $Q'$.\cite{jam05,ort07}  
For the case of 2D $d=2$ systems, a critical frustration for
$Q'$ exists since, as mentioned in Sec.~\ref{sec:intro}, it can be viewed
as an anisotropic 3D, $d=3$ system.

The stabilization of inhomogeneities which have inside smaller
inhomogeneities represents a relevant new effect to be considered in
the 2D, $\gamma=2$ phase diagram. 
These states will consist of stripes of droplets alternating with
stripes of stripes (homogeneous stripes) and will appear  in a narrow
but still finite $\overline{\phi}$ window near the first-order
morphological (homogeneous-inhomogeneous) transition.  
 
In conclusion, we have derived the Coulomb frustrated phase separation
phase diagram in 2D systems with a short-range negative
compressibility subject either to a fictitious two-dimensional
logarithmic-like Coulomb interaction and the truly three-dimensional
Coulomb interaction. Similarly to three-dimensional systems, we find
that the transition from the homogeneous phase to the inhomogeneous
phase is always first-order except for a CP. Close to the CP,
inhomogeneities are predicted to form a triangular lattice with a
subsequent transition to striped states. The transition lines
continuously evolve into the weak frustration limit. In 2D systems
subject to the three-dimensional Coulomb interaction both first order
and second order transition lines are found separated by a tricritical
point.  A proliferation
of inhomogeneities is naturally expected near the first order transition
lines. The second order transition lines make a QCP
when crossed by the change of one parameter, like doping considered
in Ref.~\onlinecite{cas95b}. Then one can expect very harmonic charge
modulations close to the QCP which evolve into more
anharmonic modulations as the distance from the QCP increases. 

 C.O. was financially supported by the {\it Stichting voor
   Fundamenteel Onderzoek der Materie} (FOM). J.L. and C.D.C. received
 support from MIUR-PRIN 2007. 

\end{section}

\end{document}